\documentclass[sigconf]{acmart}
\usepackage{graphicx,color}
\usepackage{xcolor}
\usepackage{textcase}
\usepackage{setspace} 
\usepackage[linesnumbered,ruled,vlined]{algorithm2e}
\usepackage{epstopdf}
\usepackage{epsfig}
\usepackage{amsmath,amsthm}
\usepackage{multirow}
\usepackage{hhline}
\usepackage{verbatim}
\usepackage{algorithmicx}
\usepackage{lipsum}
\usepackage{algpseudocode}
\usepackage{cases}
\usepackage{enumitem}
\usepackage[tablename=TABLE]{caption}
\usepackage{hyperref}
\usepackage{setspace}
\usepackage{listings}
\usepackage{graphics}
\usepackage{graphicx,subfigure}
\usepackage{textcomp}

\copyrightyear{2023}
\acmYear{2023}
\setcopyright{acmlicensed}\acmConference[ICDCN '24]{Proceedings of the ACM International Conference on Distributed Computing and Networking}{January 4-7 2024}{Chennai, India}
\acmBooktitle{Proceedings of the ACM International Conference on Distributed Computing and Networking, January 4-7 2024, Chennai, India}
\acmPrice{15.00}
\acmDOI{10.xx/xx.xx}
\acmISBN{xx-x-xx-x}

\DeclareCaptionTextFormat{up}{\MakeTextUppercase{#1}}

\begin{document}
\title{Application and Energy-Aware Data Aggregation using Vector Synchronization in Distributed Battery-less IoT Networks}

 \author{Chetna Singhal$^*$, Subhrajit Barick, and Rishabh Sonkar}
 \email{*chetna.iitd@gmail.com}
\orcid{0000-0002-4712-8162}
 \affiliation{\institution{Indian Institute of Technology (IIT) Kharagpur}\country{}}
\renewcommand{\shortauthors}{Chetna Singhal, Subhrajit Barick, \& Rishabh Sonkar}

\begin{abstract}
The battery-less Internet of Things (IoT) devices are a key element in the sustainable green initiative for the next-generation wireless networks. These battery-free devices use the ambient energy, harvested from the environment. The energy harvesting environment is dynamic and causes intermittent task execution. The harvested energy is stored in small capacitors and it is challenging to assure the application task execution.  The main goal is to provide a mechanism to aggregate the sensor data and provide a sustainable application support in the distributed battery-less IoT network. We model the distributed IoT network system consisting of many battery-free IoT sensor hardware modules and heterogeneous IoT applications that are being supported in the device-edge-cloud continuum. The applications require sensor data from a distributed set of battery-less hardware modules and there is provision of joint control over the module actuators. We propose an application-aware task and energy manager (ATEM) for the IoT devices and a vector-synchronization based data aggregator (VSDA). The ATEM is supported by device-level federated energy harvesting and system-level energy-aware heterogeneous application management. In our proposed framework the data aggregator forecasts the available power from the ambient energy harvester using long-short-term-memory (LSTM) model and sets the device profile as well as the application task rates accordingly. Our proposed scheme meets the heterogeneous application requirements with negligible overhead; reduces the data loss and packet delay; increases the hardware component availability; and makes the components available sooner as compared to the state-of-the-art. 
\end{abstract}



\maketitle
\setcitestyle{numbers,sort,compress}

\section{Introduction}

The future Internet of things (IoT) networks will comprise of distributed deployment of battery-less sensing and actuator control devices that will function using harvested energy from the environment. These tiny intermittent computing devices can remotely monitor environment/ objects in difficult to reach places or inaccessible spaces in a maintenance-free manner for long periods of time (even decades)~\cite{Hester2017,Colin2018}. The usage of these devices will greatly impact domains like healthcare, wildlife and forest conservation, consumer and industrial applications, infrastructure monitoring and management, and space exploration. The IoT user applications in scenarios like smart home, smart industry, smart infrastructure and smart city are more complex and can be supported by the edge or cloud resources. These require inputs from several IoT hardware modules that are distributed across a physical space over which the IoT user application asserts control~\cite{edge_cloud1}. 
In order to achieve a sustainable deployment of these devices for the above mentioned use-cases, they need suitable software interfaces that are based on the specifications of task-based energy requirements~\cite{Colin2018}.

The low-power and battery-free distributed IoT deployment requires efficient and flexible implementation of application~\cite{Mottola2019}. Furthermore, application-specific environmental and programming abstractions help conceal hidden anomalies in intermittent executions of battery-less IoT~\cite{Maioli21}. However, an efficient data aggregation framework is essential to provide an assured rate of sensor-data acquisition to support a heterogeneous IoT application in a device-edge-cloud continuum system. To facilitate this we consider the application and device state based task rates with underlying task dependencies.

\begin{figure}[!htb]
\centering
\vspace{-0.15in}
\includegraphics[width=3.2in]{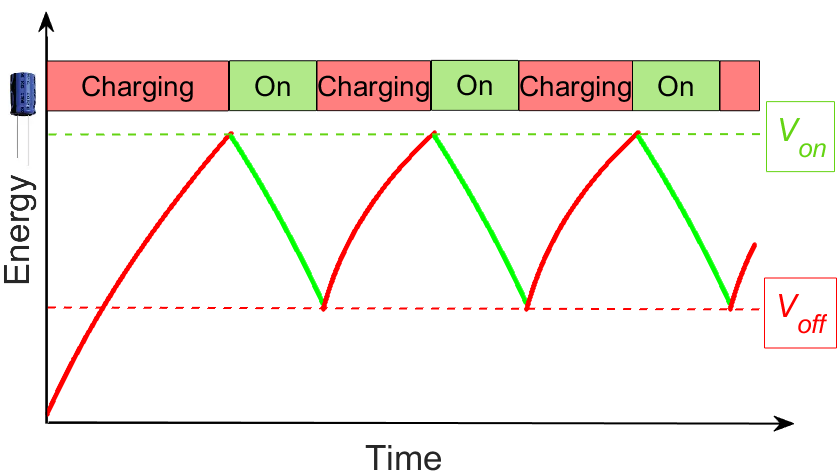}
\caption{Intermittent computing example with periodic energy buffer {\em{charging}} and IoT device {\em{on}} states.}
\vspace{-0.2cm}
\label{f:icex}
\end{figure}
The battery-less IoT devices use small capacitors that charge by harvesting energy and discharge when powering computing tasks on IoT hardware modules. The intermittent computing periodic cycles due to the charging and discharging of the capacitor is shown in Fig. \ref{f:icex}. The energy harvesting mechanism has extensive spatial and temporal unpredictability. This causes variations in the charge-discharge voltage characteristics of the capacitor~\cite{Saad2020Trans}. Such limited and varying availability of ambient energy causes the batteryless device to oscillate between task-execution and recharging phases~\cite{Saad2020EWSN}. Hence, the application support is constrained and not guaranteed in such an intermittent computing system. This is even more challenging in the distributed battery-less IoT device framework where the application is facilitated by the distributed sensor deployment in the device-edge-cloud continuum. Therefore, it is necessary to develop mechanisms that will mitigate the device state uncertainity and provides a sustainable application support in distributed batteryless-IoT device network. 

The distributed IoT network needs data aggregation to facilitate IoT user applications~\cite{data_aggregator}. The data aggregator~\cite{ti_rtos_wsn} in a battery-less IoT or low power wireless sensor network requires energy harvesting aware protocol to perform efficient data collection~\cite{data_aggregator}. The communication tasks need to be scheduled in accordance with the availability of radio interface on the battery-less IoT device based on the amount of energy harvested from the environment. Cloud computing and end-user application based data processing are integral for taking control decisions in resource constrained battery-less IoT device networks. Data aggregator facilitates this by relaying sensor and control information between the resource constrained IoT device and the user application over the network.

State Vector Synchronization is an application synchronization protocol in distributed information-centric network~\cite{VS1, VS2}. It is a simple, lightweight protocol that
synchronizes content across distributed applications by allowing periodic and event-triggered notifications. It offers sharing of data
across network participants and allows participating nodes to detect state change by comparing state vectors with the local state. The simultaneous state change updates is thereby possible with minimum  delay. In the distributed battery-less IoT network there is uncertainity associated with the active state of the sensor and radio modules. Hence, vector synchronization can be used for such a distributed network for sensor data exchange and data aggregation. Since long-short-term-memory (LSTM) has been used for forecasting time-series data \cite{access_singhal} like remaining battery life~\cite{lstm1}, we can also use LSTM model for forecasting harvester power. We propose using the LSTM based IoT device state estimator at the data aggregator, accordingly scheduling the sensor data reception and correspondingly setting the vector synchronization state.

In this paper we propose an application aware task and energy manager (ATEM) and vector-synchronization based data aggregation (VSDA) framework for intermittent computing devices. The VSDA is facilitated by LSTM-based harvester power forecasting to predict IoT device state. 
 The ATEM selects the application profile (like normal or low-power) and corresponding task rates based on application specifications. The application tasks (like sense and transmit/receive) are then scheduled and correspondingly the peripherals are controlled in the intermittent computing device. The energy management unit controls the federated energy harvesting that considers the upcoming sequence of tasks while dynamically charging the federated capacitors. Overall, this provides a sustainable application support for a distributed battery-less IoT network in the device-edge-cloud continuum. 

The ATEM scheme increases the IoT hardware component availability by at least 15.28\% and makes these components at least sooner by 22.4s than the state-of-the-art. The VSDA decreases the data loss and packet delay by 99.04\% and 94.96\%, respectively, as compared to the state-of-the-art. 

The rest of the paper is organized as follows. Section \ref{s:rel} discusses related works. Section \ref{s:sys} describes the proposed system model consisting of data aggregator and distributed IoT network to facilitate heterogeneous IoT applications in the device-edge-cloud continuum. Section \ref{s:sol} presents the proposed functional components and solution. Section \ref{s:res} provides details on the evaluation framework and presents the key performance results. Finally, Section \ref{s:conclusions} draws our conclusions.
\section{Related works}\label{s:rel}

There has been a rapid increase in the use of embedded systems and IoT. These embedded devices have diverse requirements depending upon the application domain in which they are deployed~\cite{secure2}. The battery-less IoT harvests energy from discontinuous and intermittent ambient energy sources. This drives the on-board intermittent computing, characterized by frequent transitions between the charging, computing, and non-powered states~\cite{secure2}. There is a dynamic variation between the energy buffer `{\em{charging}}' and battery-less IoT device `{\em{on}}' states resulting in  intermittent computing, as shown in Fig. \ref{f:icex}. The device energy is proportional to the system voltage and frequency~\cite{T1}. The application-specific performance demands, such as, task deadlines, can be met while minimizing the power by dynamically scaling the system voltage and frequency. Dynamic voltage scaling (DVS) is one such method that maintains the required performance level by adjusting the system supply voltage to a suitable minimum~\cite{T4}. The dynamic frequency scaling (DFS) is another method that meets the required performance by modulating the system clock frequency ~\cite{T1}. Dynamic voltage and frequency scaling (DVFS) is a combination of the above two methods and it adjusts both the voltage and frequency to maximize the system efficiency~\cite{T0,T8}. We use the intermittent computing system hardware emulator to benchmark our proposed solution in terms of its energy-consumption overhead. We implement the solution on the emulator using the MSP430x microcontroller that employs DVFS for its supported active and low-power modes. 

The ambient energy source output can extensively vary, both temporally and spatially. Hence, the energy harvesting systems incorporate large external energy buffers (such as rechargeable batteries or supercapacitors) to sustain computation. In energy-neutral operation unlimited operation is provided by ensuring that the stored energy never completely depletes~\cite{T9}. There is an attempt to balance the long-term energy consumption against the harvested energy over a period of time (e.g., a day). Energy neutrality can smooth the long-term variability in energy harvesting supplies, but requires time to charge, poses environmental issues and deteriorates in performance over time. In task-based approaches sufficient energy storage is used to execute small tasks~\cite{T9}. In our proposed ATEM-VSDA scheme, we use energy management and application-aware task execution in battery-less IoT devices.

Battery-less IoT devices can also employ transient computing and a power-neutral operation. The device operation can be directly supported from energy harvesting without any energy storage. This requires matching the instantaneous power consumption of the device to the instantaneous harvested power. It can be achieved by using control algorithms for DFS and system voltage thresholds. Here, the system performance gracefully modulates in response to the incoming power~\cite{T8}. Additionally a software based maximum power point (MPP) tracking can also facilitate the power-neutral operation. It can adapt the power consumption of the system that is operating at an efficient operating voltage and maximizing forward application execution without adding any external tracking or control units~\cite{T9,T33}. We propose a solution for data aggregation in distributed intermittent computing IoT battery-less device network. In our proposed scheme, we implement the application aware task and energy manager within the MSP430x microcontroller that manages the system performance dynamically without the need of any external functional unit. 

Ambient energy source like kinetic energy can be effectively used in low-power wearables~\cite{T79}. Such a device can harvest energy during different human activities. It could acquire more than 2000 images for an hour (7  power saving modes) while being in one of the modes: sleep, acquire, store, and near-field-communication (wireless) send~\cite{T79}. In such an implementation voltage-current characteristics of the kinetic energy harvesting transducer is used to find the optimal operating point, i.e. MPP, dynamically~\cite{T75}. The MPP sampling rate and harvesting efficiency facilitates dynamic MPP tracking~\cite{T75}. In this work, we evaluate our proposed ATEM-VSDA framework with solar energy harvesting and diverse set of applications. 

Sensor nodes with energy harvesting modules store energy in a buffer and periodically sense a random field (such as temperature, humidity) that generates a packet~\cite{T69,T78}.
These packets are stored in a queue and transmitted using the energy available in the buffer at that time. Stability needs to be ensured for the sensor data queue while choosing the highest data rate for transmission under varying channel conditions~\cite{T69,T78}. The packet transmission throughput can be optimized while minimizing the mean delay using a greedy policy in the low SNR regime~\cite{T78}. Energy management policies play an important role in such intermittent computing system to support applications and sensor data transmission. Such policies are more efficient when based on the current and past harvested energy observations rather than future predictions~\cite{T69}. 

The sporadic energy availability in intermittent computing systems makes the real-time task scheduling difficult~\cite{islam2020, karimi2021}. However, the schedulability can be improved by dynamically scheduling the computational and energy harvesting tasks, as done in Celebi~\cite{islam2020}. The schedulability performance with Celebi is further enhanced along with  periodic execution of sensing tasks by using a real-time periodic task scheduling framework~\cite{karimi2021}. The battery-less IoT devices can also partition and prioritize harvested energy into multiple isolated smaller energy buffers (capacitors). Task scheduling and different voltage requirements of the peripherals can be facilitated by these capacitors based on the application~\cite{Hester2015}. This mechanism is known as federated energy storage~\cite{Hester2015}. In our earlier work, we  have developed the application support and energy-attack mitigation frameworks for battery-less IoT~\cite{Singhal2023_icc, singhal2023_mobiwac}. In this paper, we propose application-aware task and federated energy harvesting manager, ATEM, framework that works in conjunction with the vector synchronization based data aggregator, VSDA.

An on-demand, coordinated, energy adaptive duty cycle based slotted cyclic TDMA (time division multiple access) scheme \cite{tdma_ref1} is used for efficient data collection and control information dissemination in battery-less IoT device networks. Adaptive duty-cycling \cite{tdma_ref1} allows an efficient coordinated exchange of control-information and sensor-data between the data aggregator and the battery-less IoT device. The collection of information from battery-free devices in smart-home (with RFID technology) is possible using adaptive MAC protocol, APT-MAC, for supporting applications like object identification and counting \cite{da_mac_1}. In TDMA, the number of slots in a given time cycle  and the slot as well as cycle duration are governed by the number of available IoT devices in the network that are associated with the given data aggregator \cite{tdma_ref1}. 

Distributed information centric networks synchronize the data collection using vector synchronization \cite{VS1,VS2}. Event-triggered Consensus-based Vector Synchronization protocol was designed for information collection in~\cite{vs4}. Blind synchronization without transmitter-side information is possible using state vector synchronization~\cite{vs3}. 
LSTM model can be used to forecast  battery charging/discharging time-series data~\cite{lstm1}. LSTM model has also been previously used for the battery state-of-charge estimation~\cite{lstm2}. Hence, in our proposed data-aggregator vector synchronization based scheme we use LSTM model to forecast the harvester power value and use it to predict the IoT device state. 

The following are a few key contributions of this work:
\vspace{-0.02in}
\begin{itemize}
\item LSTM-based IoT device state estimate and corresponding scheduling of beacon packets at the data aggregator.
\item Heterogeneous IoT application and distributed IoT network modeling.
\item Vector-synchronization based sensor data aggregation (VSDA) for distributed battery-less IoT network.
\item Application aware task scheduling and federated energy management (ATEM) at the battery-less IoT device.
\end{itemize}
\vspace{-0.1in}
\section{System model} \label{s:sys}
We consider the set of IoT applications, denoted as  $\mathbf{A}=\{\mathcal{A}_i|1\leq i \leq A\}$. Each IoT application $\mathcal{A}_i$ executes over a set of battery-free IoT hardware modules (or devices), $\mathbf{M}_i=\{\mathcal{M}_{i,j}|1\leq j\leq M_i\}$. Each module $\mathcal{M}_{i,j}$ consists of an MCU, a transceiver, a transducer, and a set of sensors, $\mathbf{S}_{i,j}=\{\mathcal{S}_{i,j,k}|1\leq k\leq S_{i,j}\}$.
The rate of data acquisition by sensors $\mathcal{S}_{i,j,k}$ is denoted as $\mathcal{R}_{i,j,k}$. The system-level sensor-data acquisition rate is maintained in a set, $\mathbf{R}=\{ \mathcal{R}_{i,j,k}|\forall i,j,k\}$, at the data aggregator.
The Fig. \ref{f:scenario} shows a sample scenario with three IoT applications, i.e., $A=3$, with corresponding number of distributed IoT modules as $M_1=2, M_2=4,$ and $M_3=5$, in the device-edge-cloud continuum. The data-aggregator is always-on, powered, and not battery-less. The device state is estimated at the data aggregator using LSTM forecasting and the vector synchronization state is maintained and updated based on application specifications and successful sensor data acquisition. The IoT applications are heterogeneous and are facilitated by processing the sensor data at the edge/cloud.
\vspace{-0.05in}
\subsection*{Communication framework and Packet schedule}
\begin{figure}[!htb]
\centering
\includegraphics[width=3.45in]{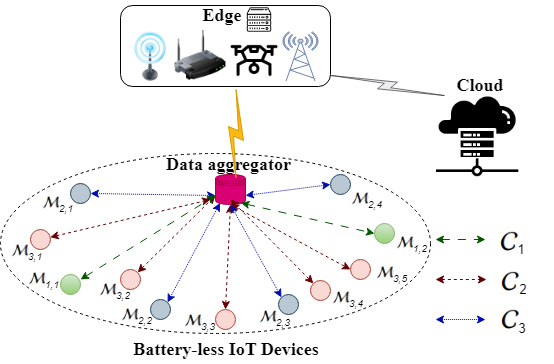}
\caption{Device-edge-cloud continuum scenario with distributed IoT device network.}
\vspace{-0.2cm}
\label{f:scenario}
\end{figure}

The data aggregator and the IoT devices (hardware modules) communicate using Bluetooth low energy (BLE) that has 40 distinct radio channels, each with 2MHz bandwidth \cite{BLE}. Each IoT application $\mathcal{A}_i$ uses a distinct radio channel $\mathcal{C}_i$. This limits the number of applications per data aggregator in the system to 40, i.e., $A=40$. In the sample scenario shown in Fig. \ref{f:scenario}, the sensor data of the three applications is sent over BLE channels $C_1, C_2,$ and $C_3$.

\subsection*{Message type}
We define the following three types of messages that are exchanged between the  battery-free IoT hardware module and the data aggregator to facilitate the proposed vector synchronization based data aggregation for the provision of heterogeneous IoT applications.
\lstset{
  language=C,                
  stepnumber=1,                   
  numbersep=5pt,                  
  backgroundcolor=\color{white},  
  showspaces=false,               
  showstringspaces=false,         
  showtabs=false,                 
  tabsize=2,                      
  captionpos=b,                   
  breaklines=true,                
  breakatwhitespace=true,         
}
\begin{enumerate}
    \item[1. ] 
\begin{lstlisting}
sensor_data{
    float *sensor;
};
\end{lstlisting} 
    This message includes the sensors' reading from the hardware module.
    \item[2. ] 
\begin{lstlisting}
rate_control{
    int *rate_current;
    int *rate_new;
};
\end{lstlisting} 
    This message exerts the rate control of sensor data acquisition at the IoT hardware module. The current and new rate values for the module sensors is given by \texttt{rate\_current[]} and \texttt{rate\_new[]}, respectively. 
    \item[3. ] 
\begin{lstlisting}
app_synch{
    int *synch_vector_current;
    int *synch_vector_new;
};
\end{lstlisting} 
    This message facilitates state vector synchronization between the IoT modules associated with an application and the data aggregator. 
    \item[4. ] 
\begin{lstlisting}
actuator_control{
    int state; // on=1, off=0
    float degree_control;
};
\end{lstlisting}
    This message exerts actuator control by setting the \texttt{state} value to `1' or `0' for turning the actuator `on' or `off'.
\end{enumerate}

\subsection*{Packet type and Packet schedule}
The packets exchanged between the data aggregator and the IoT modules are categorized as below:
\begin{enumerate}
    \item[1. ] Beacon packet is broadcast from the data aggregator to the sensor nodes. It contains the \texttt{rate\_control}, \texttt{app\_synch}, and \texttt{actuator\_control} messages. In \texttt{app\_synch}, we denote the \texttt{synchronization\_vector\_current} as $V_i=\{\mathcal{V}_{i,j}|\forall i,j\}$ and \texttt{synchronization\_vector\_new} is denoted as $\widehat{V}_i=\{\widehat{\mathcal{V}_{i,j}}|\forall i,j\}$. Here, $\mathcal{V}_{i,j}$ denotes the current count of sensor data readings successfully received from the IoT module $M_{i,j}$ in the ongoing time period of duration $T$.   $\widehat{\mathcal{V}_{i,j}}$ denotes the desired new count of sensor data readings to be received from the IoT module $M_{i,j}$ immediately following the current \texttt{Beacon} packet.
    \item[2. ] Sensor data packet is sent from the IoT module to the data aggregator. It contains the \texttt{sensor\_data} message.
\end{enumerate}

The beacon packet for each application $\mathcal{A}$ is broadcast on the BLE channel $\mathcal{C}$ periodically once in every $\tau$ time duration. This packet indicates the sense (and data acquisition) task rate of the associated distributed  IoT hardware modules. It also contains the current and new state synchronization vector indicating the desired sensor node to send the sensor information. Furthermore based on the edge/cloud based processing of sensor data obtained at the data aggregator and the user preferences the actuator control is asserted by the beacon.

Since, there is no guarantee on whether the battery-less IoT node is \texttt{on} to successfully receive the beacon packet, the same packet is resent a few times till the corresponding \texttt{sensor\_data} is successfully received or the reattempt count exceeds the predefined limit \texttt{reattempt\_count}.




\subsection*{Device state, Task state, and Task dependency}

\begin{figure}[!htb]
\centering
\vspace{-0.15in}
\includegraphics[width=3.45in]{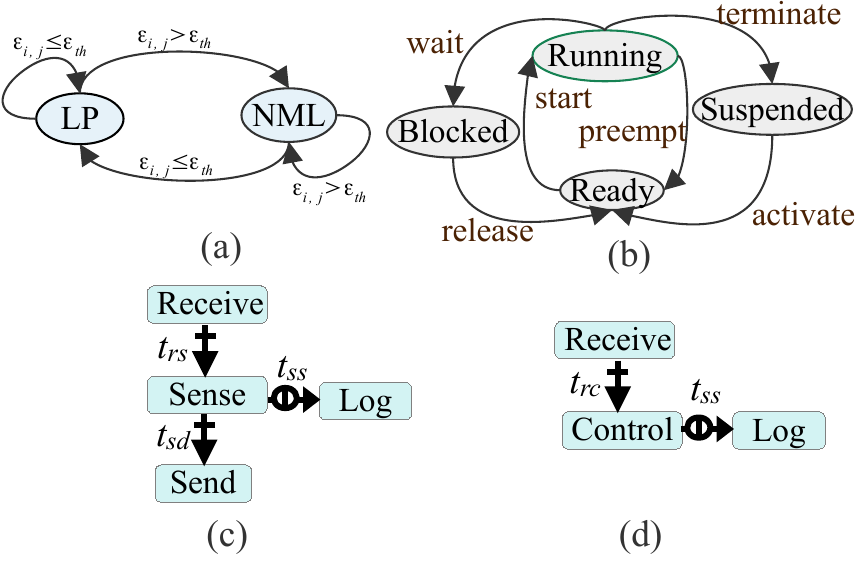}
\caption{Battery-less IoT network (a) device state transition diagram (b) task states transition diagram (c) task dependency for sensing data transmission (d) task dependency for actuator control.}
\vspace{-0.2cm}
\label{f:task_state}
\end{figure}
The device state, $\mathcal{D}_{i,j}$, of device $\mathcal{M}_{i,j}$ is determined by the harvested energy available at the device, i.e., $\mathcal{E}_{i,j}$. The system-level current device-state is maintained at the data aggregator in a set, $\mathbf{D}=\{\mathcal{D}_{i,j}|\forall i,j,k\}$. The estimated device-state for the next slot at the data aggregator correspondingly is maintained as the set $\widehat{\mathbf{D}}=\{\widehat{\mathcal{D}_{i,j}}|\forall i,j,k\}$.  The device state transition diagram is shown in Fig. \ref{f:task_state}(a). Based on the energy threshold $\mathcal{E}_{th}$, the state is set to \texttt{low\_power} (LP) if $\mathcal{E}_{i,j}\leq\mathcal{E}_{th}$ and is set to \texttt{normal} (NML), otherwise. 

The IoT application specifies the battery-less device task list and the corresponding device state based task rates. The task states are controlled by the task manager on-board the IoT hardware module. The task state transition diagram is shown in Fig. \ref{f:task_state}(b). The peripherals and resources that need energy in an intermittent computing system are: MCU (computing), sensors (sensing), and radio (transmit-receive)~\cite{Sahoo2021}. Each IoT device task can be in one of the four states shown in Fig. \ref{f:task_state}(b). Active tasks are set to be in \texttt{ready} state and the one being executed is in \texttt{running} state. Task dependency sets a task to be in \texttt{blocked} state if it is dependent on another one that is yet to be executed. 

The tasks and their timing dependency for sensing data acquisition and actuator control are shown in Figs.  \ref{f:task_state}(c) and  \ref{f:task_state}(d), respectively. The IoT device performs the beacon \texttt{Receive} task and performs the \texttt{Sense} task thereafter if the vector synchronization state in the beacon requires a sensor reading from the IoT device. Thereafter the sensor data is transmitted to the data aggregator by the \texttt{Send} task. In case the actuator control is required as per the beacon, the IoT device performs beacon \texttt{Receive} followed by the actuator \texttt{Control} task. In both the above cases, the \texttt{Log} task is performed if sufficient energy is available in the capacitor after performing the \texttt{Sense} and \texttt{Control} tasks. The task manager sets the state of the tasks and controls the IoT hardware peripherals accordingly. It schedules tasks and selects task for run-time execution.

    
 \section{Proposed Solution } \label{s:sol}
\begin{figure}[!htb]
\centering
\includegraphics[width=3.45in]{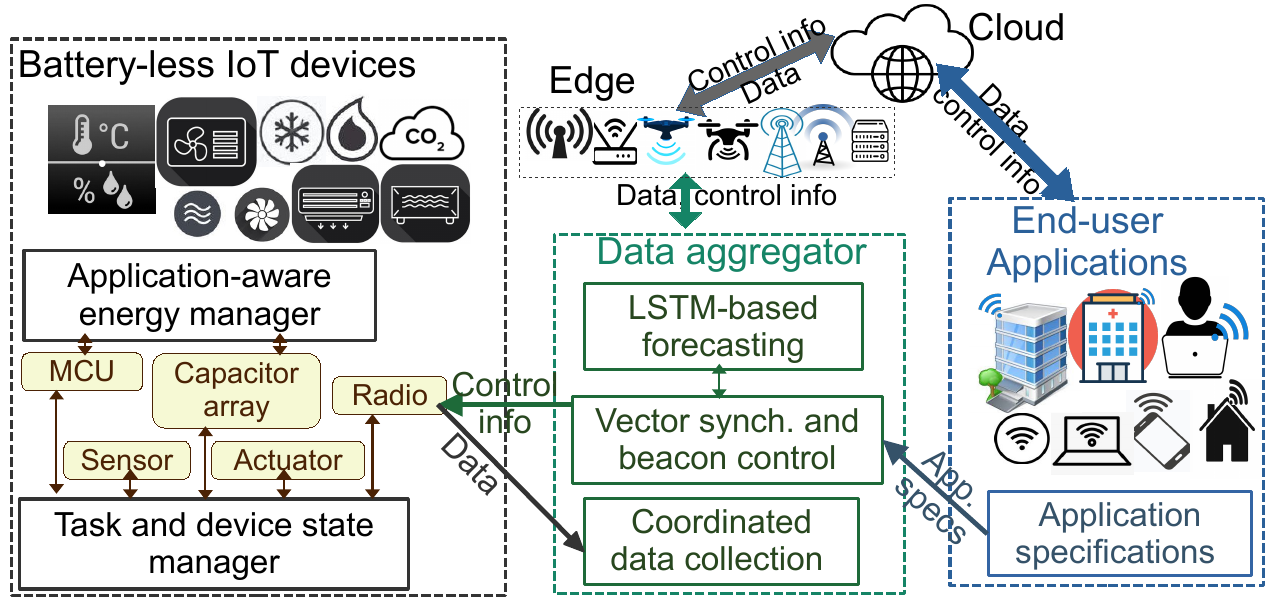}
\caption{Proposed functional blocks for the data aggregation and application provision in a distributed IoT device network architecture.}
\vspace{-0.2cm}
\label{f:arch}
\end{figure}

The proposed data aggregation and IoT device task and energy manager framework in a distributed battery-less IoT device network architecture is shown in Fig. \ref{f:arch}. The battery-less IoT device has an application-aware energy manager and a task and device state manager. These combinedly constitute the ATEM functional block that manages the hardware components (peripherals) of the IoT module. The data aggregator performs LSTM-based forecasting of IoT device harvester power as well as vector synchronization and beacon control. The coordinated data collection at the data aggregator is implemented using packet scheduling. Overall, these constitute the VSDA functional block at the data aggregator. The end-user application provides the application specifications that includes the device state (NML or LP) based IoT module sense task rate, $\rho=\{r^{LP}_{i,j},r^{NML}_{i,j}\}$, and task dependency list. The actuator control and the sensor data processing is performed using the edge and cloud computing resource due to the limited resource (energy and computing) and capabilities of the battery-less IoT device.

\subsection*{ATEM: Application-aware Task and Energy Manager}
The Algorithm \ref{atem} shows ATEM operation through psuedocode. The inputs include the application specification ($\rho$), IoT device energy ($\mathcal{E}_{i,j}$), and state vectors for synchronization ($V_i$,$\widehat{V}_i$). The output of the algorithm is the device state $\mathcal{D}_{i,j}$, sensor data acquisition rate ($\mathbf{R}$), and energy to store in the capacitor buffers ($E_1, E_2$). 

We consider an array of capacitors for federating the energy storage~\cite{Hester2015}. The lightweight task like \texttt{sense}, is performed using a smaller capacitor that charges more quickly, whereas more power-intensive tasks like \texttt{radio-transmit}, \texttt{radio-receive} is performed using the energy stored in a larger capacitor. 

The \texttt{Task\_Manager} sets the device state and task state based on the energy availability in buffers ($\mathcal{E}_{i,j}$) and task execution order. We have used two small isolated capacitors for the federated energy storage. The device state $\mathcal{D}_{i,j}$ is set to LP if $\mathcal{E}_{i,j}\leq \mathcal{E}_{th}$ and NML, otherwise. The sensor data acquisition rate is set according to $\rho$ and device state. The energy stored in first capacitor is used for MCU and sensing. The energy stored in second capacitor is  used for radio unit. The transition of a `ready' to `running' state, for a task, is performed when the energy available in the corresponding energy buffer ($E_1$ or $E_2$) is greater than that required to execute the task ($E_{sense}$, $E_{receive}, $ or $E_{transmit}$). At each IoT device the task states are managed based on the application specifications $\rho$ and vector synchronization information $V_i$,$\widehat{V}_i$ for the given device state (\texttt{NML}, \texttt{LP}). Furthermore the application requirements $\rho$ and the task (as well as device) state is used to manage the energy storage in the federated capacitor bank.

The ATEM framework consists of \texttt{Energy\_manager} that implements the charging controller using the energy harvester output $\mathcal{E}_{i,j}$. The MCU turns on using the energy stored in the first stage capacitor i.e., $E_1$. The MCU initiates the proportional charging of the isolated federated energy buffers (small capacitors). 
The voltage equations of the energy harvesting circuit is obtained by assuming power from the energy source as $P(t)$ at time $t$. We consider a parallel resistor-capacitor circuit with the equivalent storage capacitor, $c_i$ in parallel to the resistor equivalent with the rest of the circuit's equivalent resistor, $r_p$. Hence the voltage of the capacitor, $c_i$ at time instance $t$ is given as:
\begin{equation}
    v_i(t)=\sqrt{P(t)\cdot r_p - e^\frac{-2t}{c_i\cdot r_p}\cdot \left( P\cdot r_p - v_0^2\right)}
\end{equation}
where $v_0$ is the capacitor voltage at $t=0$.

The energy harvesting power observed during slot $n$ from source to buffer $i$ is $P_i(n)$, which is governed by the proposed ATEM control. The energy level in buffer $i$ at the beginning of slot $n$ is $E_i(n)$. 
The energy level in buffer $i$ at the beginning of the next time slot is given as:
\begin{equation}
    E_i(n+1)=(1-\sigma_i)\cdot E_i(n)+\eta_i\cdot P_i(n)\cdot t 
\end{equation}
where, the duration of slot $n$ is $t$, the proportion of energy budget sourced by buffer $i$ is $\sigma_i$, and the efficiency of energy buffer $i$ is $\eta_i$.
The energy management module dynamically charges the capacitors in order to sustain support of the active task at the earliest. 

The energy in the buffers that supports execution of \texttt{Sense} and [\texttt{Transmit}, \texttt{Receive}] tasks is $E_1$ and $E_2$. The \texttt{Energy\_Manager} function proportionally stores the harvested energy in isolated buffers (small capacitors) based on the state of the task it supports. If the task is `ready' then a higher proportion of energy, $\Lambda$, is stored in the corresponding buffer, else a lower proportion of energy, $\lambda$, is stored.



\subsection*{VSDA: Vector Synchronization based Data Aggregator}
The Algorithm \ref{vsda} shows VSDA operation through psuedocode. The inputs include the application specification $\rho$, time period $T$, IoT device state $\mathbf{D}$, and state vector for synchronization $V_i$. The output of the algorithm is the beacon period, vector synchronization current and new information, and the \texttt{beacon}. 
The Data aggregator predicts the battery-less device state using LSTM model (\texttt{lstm}) with periodic updates with the actual device state. The LSTM model forecasts the harvested power. The rate of beacon packet transmission and its contents (scheduled sensor data request) is decided based on the the predicted device states.

The system objective is to meet the  sensor data collection rate requirement of the application ($\mathcal{R}_{i,j,k}$). This is achieved by selecting the beacon period ($\tau$), given the proportion of estimates of the device states that are correct is $\alpha$. In our proposed solution this governs the scheduling order of the sensor data collection from the distributed IoT sensor modules in a synchronized manner by the data aggregator. The system objective and corresponding beacon period is given as: 
\begin{align}
    \alpha\cdot R(\tau)- & \sum\limits_{j=1}^{M_i}\sum\limits_{k=1}^{S_j}\mathcal{R}_{i,j,k}\geq 0, \,\, \forall \,\, i \in [1\ldots A]\\
   &  R(\tau)=\frac{T}{\tau}\\
   & \tau\leq\alpha \cdot \frac{T}{\sum\limits_{k=1}^{S_j}\mathcal{R}_{i,j,k}} \label{e:beacon}
\end{align}

\begin{algorithm}[!htb]
\small
\caption{ ATEM: Application-aware Task and Energy Manager}
\label{atem}
\begin{algorithmic}
\State \textbf{Input:} $\rho$, $\mathcal{E}_{i,j}$, $\mathcal{E}_{th}$, $V_i$,$\widehat{V}_i$\\
 \SetKwFunction{FMain}{Main}
  \SetKwFunction{FSum}{\!\!}
  \SetKwFunction{FSub}{\!\!}
  \SetKwProg{Fn}{Task\_Manager}{:}{}
  \Fn{\FSum{$\rho$, $\mathcal{E}_{i,j}$, $\mathcal{E}_{th}$,$V_i$,$\widehat{V}_i$}}{
\begin{enumerate}
 \item [1)] {\bf Select device state $\mathcal{D}_{i,j}$:} 
  \end{enumerate}
 \If{$\mathcal{E}_{i,j}\leq \mathcal{E}_{th}$}{$\mathcal{D}_{i,j}$=LP}
\Else{$\mathcal{D}_{i,j}$=NML}
  \begin{enumerate}
\item [2)] {\bf Assign sensor data acquisition rate:} 
\end{enumerate}
\For{$i=1$ to $A$}{
\For{$j=1$ to $M_i$}{
\For{$k=1$ to $S_{i,j}$}{
\texttt{state}=$\mathcal{D}_{i,j}$\newline
${}$\hspace{2em}$\mathcal{R}_{i,j,k}=r^{\texttt{state}}_{i,j}$
}}}
 \begin{enumerate}
     \item [3)] {\bf Set task state:}
\end{enumerate}
\If{$\widehat{\mathcal{V}_{i,j}}>\mathcal{V}_{i,j}$}{ \lIf{$\mathcal{E}_{i,j}>E_{sense}$ }{\newline
Set \texttt{Sense} task state to \texttt{running}}
\lElse{Set \texttt{Sense} task state to \texttt{ready}}
}
${}$\hspace{-1em}\Else{Set \texttt{Sense} task to \texttt{blocked}}
${}$\hspace{-1em}\lIf{\texttt{Sense} is \texttt{running}}{\newline 
${}$\hspace{2.5em}Set \texttt{Transmit} task to \texttt{ready}}
${}$\hspace{-1em}\lElse{\newline ${}$\hspace{2.5em}Set \texttt{Transmit} task to \texttt{blocked}}
${}$\hspace{-1em}\lIf{\texttt{Sense} and \texttt{Transmit} tasks  is \texttt{suspended}}{\newline ${}$\hspace{2.5em}Set \texttt{Receive} task to \texttt{ready}}
${}$\hspace{-1em}\lElse{Set \texttt{Receive} to \texttt{blocked}}
${}$\hspace{-1em}\lIf{\texttt{Transmit} is \texttt{ready} and $\mathcal{E}_{i,j}>E_{transmit}$}{\newline ${}$\hspace{2.5em}Set \texttt{Transmit} to \texttt{running}}
${}$\hspace{-1em}\lIf{\texttt{Receive} is \texttt{ready} and $\mathcal{E}_{i,j}>E_{receive}$}{\newline ${}$\hspace{2.5em}Set \texttt{Receive} to \texttt{running}}    
 \KwRet $\mathcal{D}_{i,j}$, $\mathbf{R}$
  }
\noindent\SetKwProg{Fn}{Energy\_Manager}{:}{}
  \Fn{\FSub{$\mathcal{E}_{i,j}$}}{
  \begin{enumerate}
     \item [1)] {\bf Proportion of harvested energy to capacitor}
\end{enumerate}
 ${}$\hspace{1em}
\lIf{\texttt{Sense} is \texttt{ready} or \texttt{running}}{\newline${}$\hspace{2.5em}{\bf Set} $E_1=\Lambda\cdot \mathcal{E}_{i,j}$
\newline${}$\hspace{2.5em}{\bf Set} $E_2=\lambda\cdot \mathcal{E}_{i,j}$
}
  \lElse{\newline${}$\hspace{2.5em}{\bf Set} $E_1=\lambda\cdot \mathcal{E}_{i,j}$
  \newline${}$\hspace{2.5em}{\bf Set} $E_2=\Lambda\cdot \mathcal{E}_{i,j}$
  }
  \KwRet $E_1, E_2$
  }
\noindent\textbf{Output:} $\mathcal{D}_{i,j}$, $\mathbf{R}$, $E_1, E_2$
\end{algorithmic}
\end{algorithm}

\begin{algorithm}[!htb]
\small
\caption{VSDA: Vector Synchronization based Data Aggregation}
\label{vsda}
\begin{algorithmic}
\State \textbf{Input:} $\rho$, $T$, $\mathbf{D}$, $V_i$, $\mathcal{E}_{i,j}(\tau_n)$, $\forall 1\leq i\leq A$, $\mathcal{E}_{th}$\\
 \SetKwFunction{FMain}{Main}
  \SetKwFunction{FSum}{\!\!}
  \SetKwFunction{FSub}{\!\!}
  \SetKwProg{Fn}{Device\_state}{:}{}
  \Fn{\FSum{$\mathbf{D}$}}{
\begin{enumerate}
 \item [1)] {\bf Estimate energy $\mathcal{E}_{i,j}(\tau_{n+1})$ in next time slot, $\tau_{n+1}$:} 
  \end{enumerate}
 ${}$\hspace{1em} \For{$i=1$ to $A$ }
  {  
  \For{$j=1$ to $M_i$}
  {
  $(\alpha,\mathcal{E}_{i,j}(\tau_{n+1}))$=\texttt{lstm}($\mathcal{E}_{i,j}(\tau_{n})$)
  }
  }
 \begin{enumerate}
\item [2)] {\bf Estimate the IoT device state $\widehat{\mathcal{D}_{i,j}}$:} \newline 
\If{$\mathcal{E}_{i,j}(\tau_{n+1})\leq \mathcal{E}_{th}$}{$\widehat{\mathcal{D}_{i,j}}$=LP}
\Else{$\widehat{\mathcal{D}_{i,j}}$=NML}
\end{enumerate}
  \begin{enumerate}
\item [3)] {\bf Assign task rate:} 
\end{enumerate}
\For{$i=1$ to $A$}{
\For{$j=1$ to $M_i$}{
\For{$k=1$ to $S_{i,j}$}{
\texttt{state}=$\widehat{\mathcal{D}_{i,j}}$\newline
${}$\hspace{2em}$\mathcal{R}_{i,j,k}=r^{\texttt{state}}_{i,j}$
}}}

\KwRet $\mathbf{R}=\{\mathcal{R}_{i,j,k}|\forall i,j,k\}$, $\alpha$
}
\noindent\SetKwProg{Fn}{Vector\_synchronization}{:}{}
  \Fn{\FSub{$\mathbf{R}$, $T$, $\alpha$}}{
\begin{enumerate}
     \item [1)] {\bf Set beacon period $\tau$ using \eqref{e:beacon}}
\end{enumerate}
  \begin{enumerate}
     \item [2)] {\bf Set vector synchronization state, $\widehat{V}_i$:}
\end{enumerate}
${}$\hspace{0.2em}\For{$j=1$ to $M_i$}{
  ${}$\hspace{0.2em}\texttt{flag}=0\\
  ${}$\hspace{2em}\If{$\mathcal{V}_{i,j}<\mathcal{R}_{i,j,k}$ and \texttt{flag}==0}{  ${}$\hspace{-1em}$\widehat{\mathcal{V}_{i,j}}=\mathcal{V}_{i,j}+1$\\
  ${}$\hspace{1em}\texttt{flag}=1
  }
 }
\begin{enumerate}
     \item [3)] {\bf Packet schedule and vector state update:}
\end{enumerate}
  ${}$\hspace{0.2em} Schedule \texttt{Beacon} every $\tau$ s.\\
  \If{$\mathcal{M}_{i,j}$ \texttt{sensor\_data} is \texttt{received}}{
 $\mathcal{V}_{i,j}=\widehat{\mathcal{V}_{i,j}}$
  }
  \If{no \texttt{sensor\_data} is \texttt{received} during $\tau$}{
  Repeat \texttt{Beacon} for maximum \texttt{reattempt\_count} times
  }
  \KwRet $V_i$
  }
\noindent\textbf{Output:} $\tau$,  $\widehat{V}_i, \forall 1\leq i \leq A$, $\texttt{Beacon}$
\end{algorithmic}
\end{algorithm}


For a given application, the order of sensor information collection from the battery-less hardware modules (devices) is synchronized by the data aggregator by means of the vector sent in the beacon packet. This order is governed by the device states' and the task rates of all the IoT modules facilitating a given application. 

\subsection*{Data aggregation example}
\begin{figure}[!htb]
\centering
\vspace{-0.15in}
\includegraphics[width=3.45in]{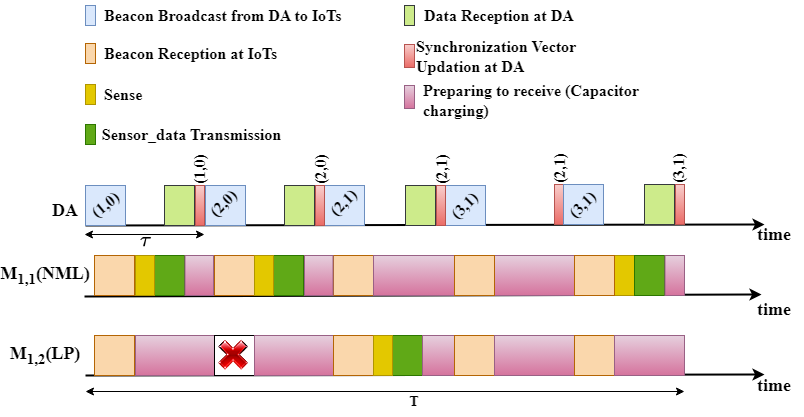}
\caption{Beacon and sensor data packet scheduling for IoT devices.}
\vspace{-0.2cm}
\label{f:timing_dia}
\end{figure}
Fig. \ref{f:timing_dia} shows an example with two IoT modules $\mathcal{M}_{1,1}$ in NML mode and $\mathcal{M}_{1,2}$ in LP mode for application $\mathcal{A}_1$. The application specification has rate requirement of [$r_{1,1}^{NML}=3, r_{1,2}^{LP}=1$] in this example. The \texttt{beacon} packet is broadcast periodically after $\tau$ time duration that is received by active IoT nodes (i.e. radio is on). Initially the data aggregator (DA) beacon has the parameters set as, \texttt{synchronization\_vector\_current}=[0,0] as well as \texttt{synchronization\_vector\_new}=[1,0]. On receiving this beacon the IoT module $\mathcal{M}_{1,1}$ sends the \texttt{sensor\_data} and the DA updates the \texttt{synchronization\_vector\_current}=[1,0] as well as the parameter \texttt{synchronization\_vector\_new}=[2,0]. The LP module $\mathcal{M}_{1,2}$ misses the second beacon packet. Since DA predicts the device state the second beacon was targeted to collect the sensor data from $\mathcal{M}_{1,1}$. Overall in the time period $T$, the rate of (3,1) from ($\mathcal{M}_{1,1}, \mathcal{M}_{1,2}$) is successfully achieved in this example.
\section{Evaluation and Results}\label{s:res}
We consider a use case of two solar energy harvesting powered battery-less IoT devices, implemented using TI MSP430FR \cite{ti_rtos_wsn} (a low power MCU), with Bluetooth low-energy transceiver. We have used indoor and outdoor measurement datasets, EnHANTS Irradiance \cite{Gorlatova_Infocom2011,dataset1}, to characterize the solar energy harvesting power available at these IoT devices. 
 An application cycle consists of completion of tasks in the following execution order: {\bf{receive-sense-send}}, {\bf{receive-control}}. For example, the sensing task needs to be performed before the radio-send task. The sensor data is sent to the cloud, with the help of a data aggregator. The control decision is taken based on data processing or user control in the end-user application. The radio-receive provides the IoT device with the decision (from the previous cycle) and control information (from user application and data aggregator) facilitating the control task in the ongoing application cycle.

 We consider three applications with specifications as given in Table \ref{t:app}. We study the comparative system performance for the three system scenarios listed in Table \ref{t:scen}. These scenarios have different proportion of LP and NML state IoT devices in the distributed IoT network. 
 We evaluate the proposed ATEM scheme in comparison to federated harvesting (FH,~\cite{Hester2015,Hester2017}) and central (with a single large capacitor) energy buffer schemes employing celebi task scheduling~\cite{islam2020}. We also perform the comparative performance analysis of VSDA scheme with respect to APT-MAC~\cite{da_mac_1} data aggregation scheme.

\begin{table}[!tbp]
\caption{Application specifications\vspace{-2mm}}
\label{t:app}
\begin{center}
\small
\begin{tabular}{|c|c|c|c|}    
    \hline
   {\bf Application}&Modules&\multicolumn{2}{|c|}{{\bf Rate (sensor readings)}}\\
&&\multicolumn{2}{|c|}{{\bf per hour per module)}}\\\cline{3-4}
 &    & NML &  LP\\\hline
    1 & 2 & 10 & 5 \\ \hline
    2&4 & 16 & 8  \\ \hline
   3 & 5& 20 & 10   \\ \hline
\end{tabular}
\end{center}
\vspace{-4mm}
\end{table}

\begin{table}[!tbp]
\caption{Performance evaluation scenarios\vspace{-2mm}}
\label{t:scen}
\begin{center}
\small
\begin{tabular}{|c|c|c|}    
    \hline
   {\bf Scenario}&\multicolumn{2}{|c|}{{\bf Proportion of devices per mode (\%)}}\\\cline{2-3}
 &     NML &  LP\\\hline
    1 & 0 & 100 \\ \hline
    2& 50 & 50  \\ \hline
   3 & 100 & 0   \\ \hline
\end{tabular}
\end{center}
\vspace{-4mm}
\end{table}

\begin{figure}
    \centering
      \subfigure[\label{f:loss_1}]{
      \includegraphics[width=0.48\columnwidth]{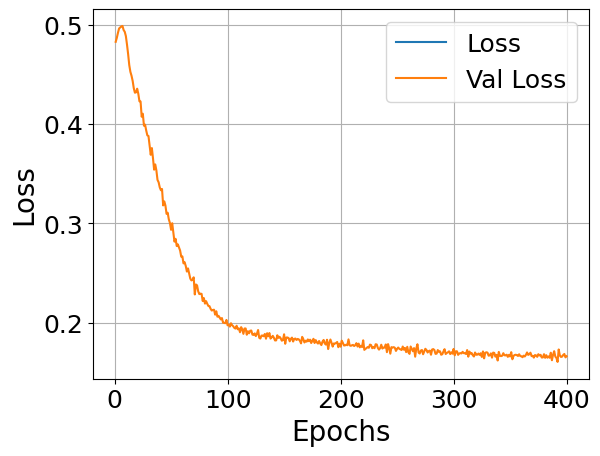}  
    }\hspace{-2mm}
         \subfigure[\label{f:loss_2}]{
         \includegraphics[width=0.48\columnwidth]{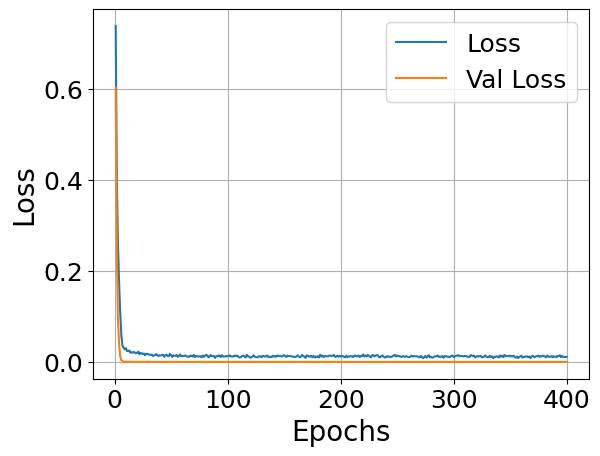}
         }
    \caption{Epoch-wise loss of LSTM model training to predict solar energy harvesting power for (a) Outdoor and (b) Mobile scenarios.}
    \label{f:lstm}
    \vspace{-5mm}
\end{figure} 

\begin{figure}
    \centering
      \subfigure[RMSE=2.473 mW\label{f:lstm_1}]{
      \includegraphics[width=0.48\columnwidth]{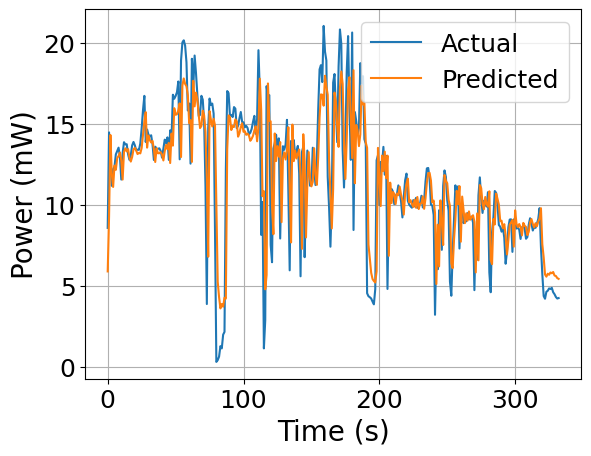}  
    }\hspace{-2mm}
         \subfigure[RMSE=0.109 mW\label{f:lstm_2}]{
         \includegraphics[width=0.48\columnwidth]{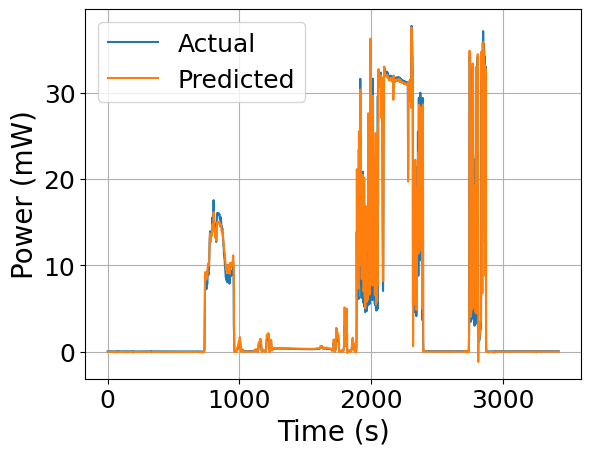}
         }
    \caption{LSTM-based prediction of solar energy harvesting power using the data from the previous 1~s duration, 100 training epochs, and 10 data values per second, for (a) Outdoor and (b) Mobile scenarios.}
    \label{f:lstm1}
    \vspace{-5mm}
\end{figure} 
The LSTM model based prediction of Solar Energy Harvesting Power is performed at the at Data Aggregator for the EnHANTS dataset. The parameters used for the LSTM model based forecaster are: 0.001 learning rate,  \texttt{Adam} optimizer, 400 epochs, and upto 10 time-series samples. Fig. \ref{f:lstm} shows the epoch-wise LSTM model training and validation loss for the \texttt{Outdoor} and \texttt{Mobile} scenarios. The loss reduces below 0.2 after 100 epochs. Fig. \ref{f:lstm1} shows the actual and predicted harvested power with time for the \texttt{Outdoor} and \texttt{Mobile} scenarios. The RMSE is less than 2.5 mW. The corresponding proportion of estimates of the device states that are correct ($\alpha$) is 0.98 and 0.99 for the \texttt{Outdoor} and \texttt{Mobile} scenarios, respectively.

\subsection*{Energy consumption overhead of IoT tasks and ATEM modules}

We evaluate the energy consumption of the functional module of the proposed solution using the intermittent computing system hardware emulator, MSPSim. We ascertain the extent of energy consumption overhead due to the execution of the proposed functional modules on the MSP430f2618 microcontroller that consumes 515 $\mu$A at 3 V (i.e. 1.545 mW power, on average) in active mode. The Dual-Mode Bluetooth low-energy transceiver, CC256x \cite{ble_cc}, is used in the IoT device for send and receive tasks. The corresponding energy consumption of the proposed functional modules are listed in Table \ref{t:mspsim_results}. We consider the  worst-case energy consumption values of the application tasks (i.e. sensor (read), send, and receive) based on the microcontroller datasheet, literature, and MSPSim-based evaluation~\cite{msp430f2618-datasheet,ea_base,MSPSim}. We have included these values of execution duration and energy consumption of the application tasks and proposed ATEM functional modules (\texttt{Application\_Manager}, \texttt{Energy\_Manager}) in the performance evaluation framework. 

\begin{table}[!htp]
\caption{Energy consumption of ATEM functional modules}
\small
\label{t:mspsim_results}
\begin{center}
\begin{tabular}{ |p{1.2in}|c|c| } 
 \hline
 \bf{Module}  &\!\!\bf{Execution\,time}\!\!& \bf{Energy}\\ \hline
Overall ATEM  & 0.582 $\mu$s & 0.782 nJ\\ \hline
 Energy\_Manager & 0.198 $\mu$s & 0.217 nJ \\ \hline
 Task\_Manager & 0.379 $\mu$s & 0.493 nJ   \\\hline
 \end{tabular}
\end{center}
\end{table}

\begin{table}[!htp]
\caption{Energy consumption of on-board IoT device tasks}
\small
\label{t:mspsim_results}
\begin{center}
\begin{tabular}{ |p{0.5in}|c|c| } 
 \hline
 \bf{Task} & \bf{Duration}&\bf{Energy} \\ \hline
Sense & 12.030 msec & 19.066 $\mu$J\\ \hline
 Send & 52.558 msec & 67.891 $\mu$J\\ \hline
 Receive & 58.483 msec & 92.931 $\mu$J \\\hline
 \end{tabular}
\end{center}
\end{table}
The execution time and energy consumption of each ATEM functional modules is lesser than all the IoT application-tasks. We observe that a single execution of the ATEM functions consumes 0.782\,nJ energy and takes 0.582 $\mu$s for execution 
and this is many order of magnitudes lesser than that of each  application task (\texttt{Sense}, \texttt{Receive}, \texttt{Send}).
\begin{figure}[!htb]
\centering
\vspace{-0.15in}
\includegraphics[width=2.5in]{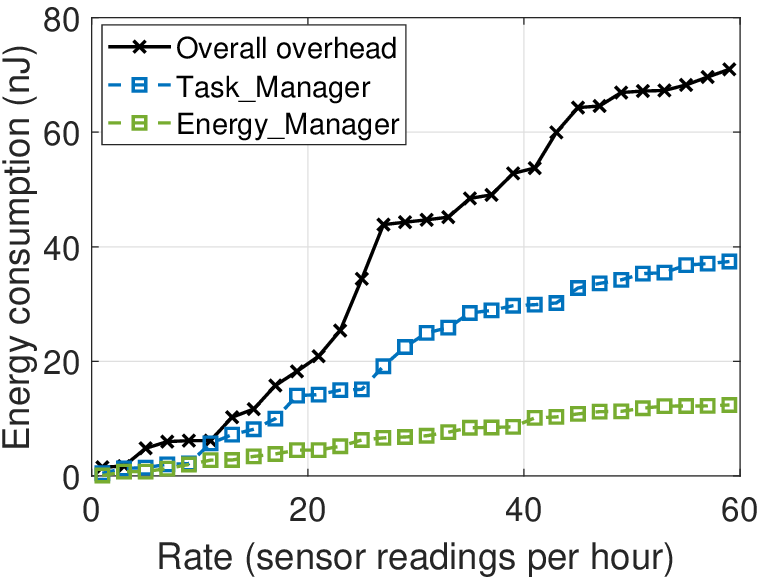}
\caption{Overhead of functional modules and overall implementation of the proposed ATEM framework.}
\vspace{-0.2cm}
\label{f:overhead}
\end{figure}
Fig. \ref{f:overhead} shows the energy consumption overhead of ATEM functional modules with increase in the sensor data acquisition rate. Even with a high rate of 50 sensor readings per hour the energy consumption is less than 65 nJ.

\subsection*{Comparative performance}
\begin{figure}
    \centering
      \subfigure[\label{f:initial}]{
      \includegraphics[width=0.485\columnwidth]{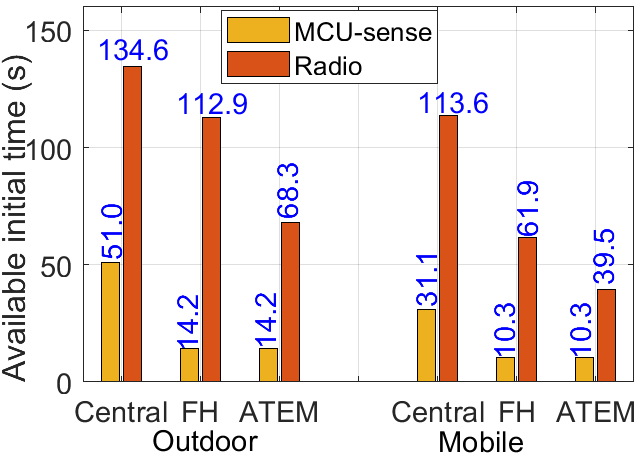}  
    }\hspace{-2mm}
         \subfigure[\label{f:available}]{
         \includegraphics[width=0.475\columnwidth]{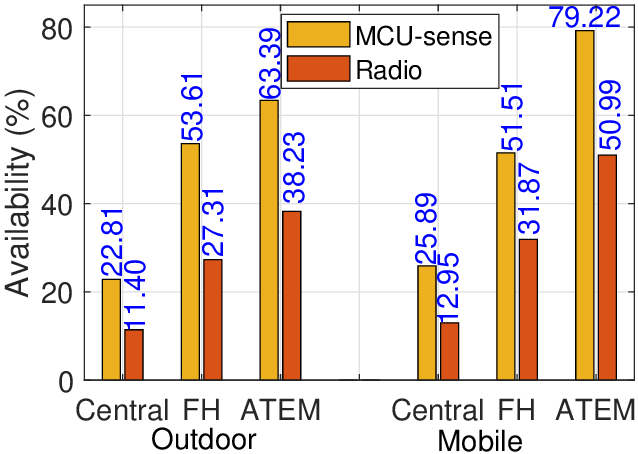}
         }
    \caption{(a) Available-initial-time and (b) Availability of hardware components with central, FH, and Application-aware task and energy manager (ATEM) schemes in Outdoor and Mobile scenarios.}
    \label{f:da_1}
    \vspace{-5mm}
\end{figure}

The ATEM comparative performance for the \texttt{Outdoor} and \texttt{Mobile} scenario, in terms of component (MCU-sense and Radio) available initial time and availability, is shown in Figs. \ref{f:available}(a) and \ref{f:available}(b), respectively. The availability of all the components (radio and MCU) with ATEM is at least 25\% higher than the Central and FH schemes. Also, the components are available at least 20~s sooner with ATEM as compared to Central and FH schemes.

\begin{figure}
    \centering
      \subfigure[\label{f:rate}]{
      \includegraphics[width=0.48\columnwidth]{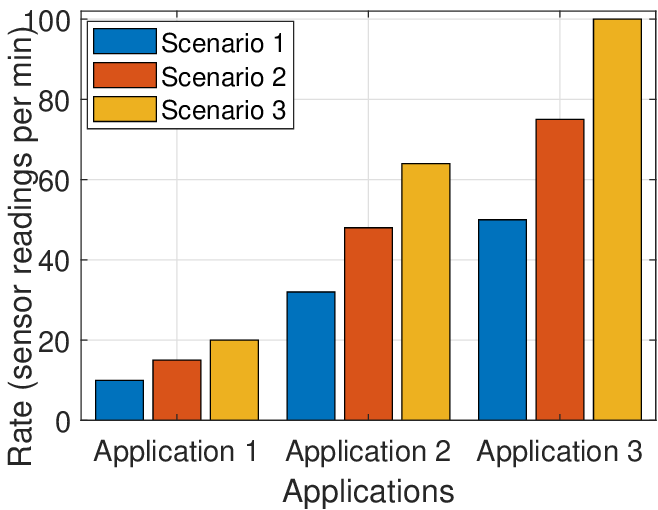}  
    }\hspace{-2mm}
         \subfigure[\label{f:tau}]{
         \includegraphics[width=0.48\columnwidth]{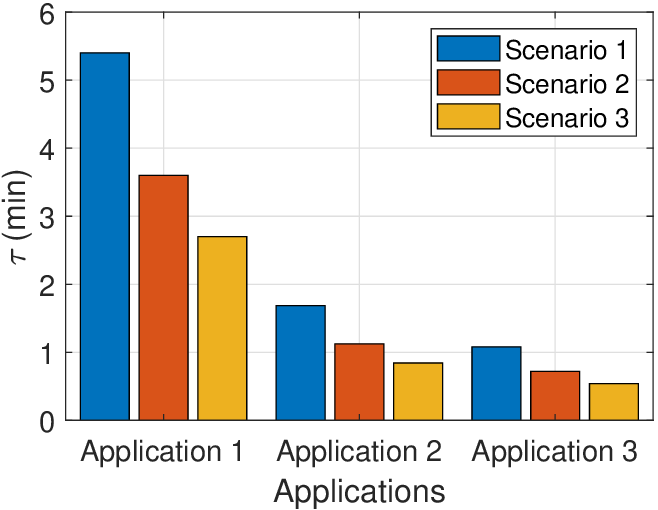}
         }
    \caption{Application-wise (a) rate and (b) beacon interval, $\tau$, for scenarios listed in Table \ref{t:scen}, with VSDA and $\alpha$ according to LSTM model devoce-state prediction performance.}
    \label{f:rate_tau}
\end{figure}

The VSDA performance for scenarios listed in Table \ref{t:scen} and applications listed in Table \ref{t:app}, in terms of the application-wise rate and beacon interval, $\tau$, is shown in Figs. \ref{f:rate_tau}(a) and \ref{f:rate_tau}(b), respectively. This corresponds to the $\alpha$ according to LSTM model prediction performance shown in Fig. \ref{f:lstm1}. The beacon interval is smaller for a high sensor data acquisition rate. The rate for each application is higher for the scenarios with more IoT nodes in the LP mode, causing lesser correct device state estimates, i.e. smaller $\alpha$.   

\begin{figure}
    \centering
      \subfigure[\label{f:rate}]{
      \includegraphics[width=0.48\columnwidth]{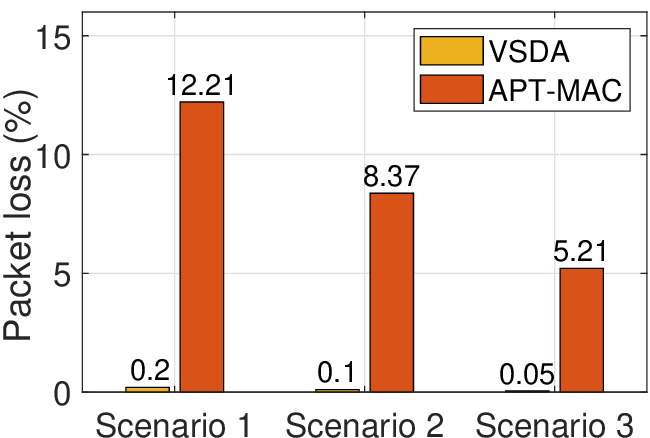}  
    }\hspace{-2mm}
         \subfigure[\label{f:tau}]{
         \includegraphics[width=0.48\columnwidth]{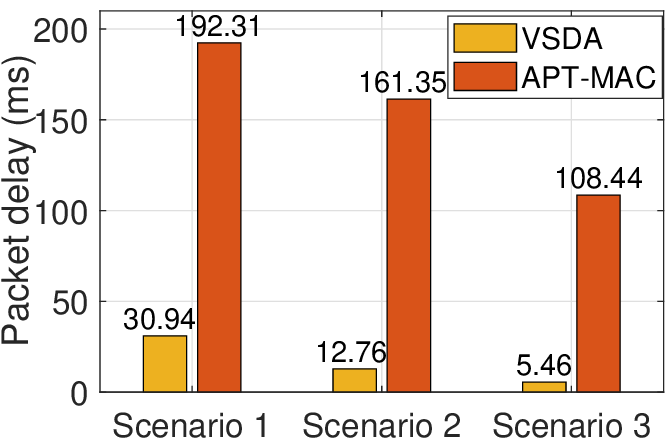}
         }
    \caption{Scenario-wise (a) Data-loss and (b) Packet delay,  for scenarios listed in Table \ref{t:scen}.}
    \label{f:pl}
\end{figure}
The data-loss and sensor-data packet delay for scenarios listed in Table \ref{t:scen} is shown in Fig. \ref{f:pl}(a) and \ref{f:pl}(b), respectively. The VSDA results in atleast 98\% lesser data loss and atleast 100 ms lesser than APT-MAC scheme. In our proposed framework, VSDA estimates the IoT device state before initiating the data acquisition and collection. Subject to device availability, the data aggregation in VSDA is application and device-state based making the packet transfer immediate and successful as compared to APT-MAC that does not estimate the IoT device state causing higher packet loss and delay. Overall, the ATEM and VSDA schemes are combinedly effective for  efficient data aggregation in distributed battery-less IoT network, as compared to the state-of-the-art.
\section{Conclusion}\label{s:conclusions}
In this paper, we have proposed an application aware task and energy manager (ATEM) for IoT devices in a distributed battery-less IoT network. We also propose a vector synchronization based data aggregator (VSDA) to facilitate heterogeneous IoT applications in device-edge-cloud continuum system. The data aggregator performs LSTM-based forecasting of the available power from the ambient energy harvester and maintains the IoT device states. The overhead of the proposed framework implementation is evaluated using battery-less hardware emulator and is found to be negligible as compared to the application tasks. We have performed comparative performance analysis for three scenarios with three heterogeneous applications and varied system conditions, with respect to the state-of-the-art schemes \texttt{central}, \texttt{FH}, and \texttt{APT-MAC}. Combinedly, the proposed framework, increases hardware component availability making them available sooner, reduces data loss and packet delay as compared to the state-of-the-art. 
In the future work, we will accommodate IoT network scalability and devise mechanisms to handle practical challenges in deployment for scenarios with extreme variability in available ambient energy.

\setcitestyle{numbers,sort,compress}
\balance
 \bibliographystyle{ACM-Reference-Format}
 \bibliography{biblio}
\end{document}